\documentclass[12pt]{article}

\usepackage{amsmath}
\usepackage{amsfonts} 
\usepackage{amssymb}
\usepackage{graphicx}
\usepackage{hyperref}
\usepackage{tensor}
\usepackage{siunitx}
\newcommand{\be}{\begin{equation}}
\newcommand{\ee}{\end{equation}}
\usepackage{xcolor}

\newcommand{\arXiv}[2]{\href{http://arxiv.org/pdf/#1}{{\tt [#2/#1]}}}
\newcommand{\arXivold}[1]{\href{http://arxiv.org/pdf/#1}{{\tt [#1]}}}

\newcommand{\beq}{\begin{equation}}  \newcommand{\eeq}{\end{equation}}
\newcommand{\bal}{\begin{aligned}}   \newcommand{\eal}{\end{aligned}}
\newcommand{\bea}{\begin{eqnarray}}  \newcommand{\eea}{\end{eqnarray}}

\def\mp{M_{\text{\tiny P}}}

\def\mpp{M^2_{\text{\tiny P}}}
\newcommand{\bh}{\text{\tiny BH}}

\usepackage{geometry}
\begin{document}
\vspace{5mm}
\vspace{0.5cm}
\vspace*{1cm}
\begin{center}

\def\thefootnote{\fnsymbol{footnote}}
\thispagestyle{empty}
{\large \bf 
 A Note on  the Swampland Distance Conjecture 
}
\\[1.5cm]
{\large   A. Kehagias$^{a}$ and A. Riotto$^{b}$}
\\[0.5cm]

\vspace{.3cm}
{\normalsize {\it  $^{a}$ Physics Division, National Technical University of Athens, \\15780 Zografou Campus, Athens, Greece}}\\

\vspace{.3cm}
{\normalsize { \it $^{b}$ Department of Theoretical Physics and Center for Astroparticle Physics (CAP)\\ 24 quai E. Ansermet, CH-1211 Geneva 4, Switzerland}}\\

\vspace{.3cm}


\end{center}

\vspace{3cm}

\hrule \vspace{0.3cm}
{\small  \noindent \textbf{Abstract} \\[0.3cm]
\noindent 
We  discuss  the Swampland Distance Conjecture in the framework of black hole thermodynamics. In particular, we consider black holes in de Sitter space and we show that the Swampland Distance Conjecture is a consequence of the fact that apparent horizons are always inside cosmic event horizons whenever they exist in the case of fast-roll inflation.  In addition, we show that the Bekenstein and the Hubble entropy bounds for the entropy in a region of spacetime lead similarly  to the same conjecture. 

\vspace{0.5cm}  \hrule
\vskip 1cm

\def\thefootnote{\arabic{footnote}}
\setcounter{footnote}{0}



\baselineskip= 17pt

\newpage 
\clearpage
\pagenumbering{arabic}

\section{Overview}
A fundamental paradigm  in  modern cosmology is that there was a period in the early universe  where the vacuum energy dominates giving rise  to a phase called inflation \cite{lrr}. During such an inflationary phase the  scale factor grows  exponentially and the spacetime geometry is quasi-de Sitter (dS). 
A small,  spatial region grows so large as to  encompass the portion of the universe we observe today, thus solving the horizon problem. At the same time during inflation
scalar quantum fluctuations are generated, which are ultimately responsible for the universe we observed today.

All inflationary models constructed so far assume an almost  dS background and therefore dS spacetime should exist in a quantum theory of gravity.
However, it has been proved notoriously difficult to construct dS spaces (stable
or even metastable)  in a fundamental theory such as string theory. This difficulty can be  attributed to a fundamental property of the theory 
itself and it has been extended to the so-called  dS Swampland Conjecture \cite{OOSV} (see \cite{BV,Palti} for  reviews). According to its original formulation,   the scalar potential $V(\phi)$ of the inflaton field $\phi$ driving inflation must satisfy the condition $|\partial_\phi V|\geq c V$ where $c$ is a constant of order unity. This conjecture was based on previous works \cite{V1,OV,g1,g2,g3} and has drawn a lot of attention (see \cite{dSC2}-\cite{Blum} for some references). It should be stressed that although there is no at the moment general proof of the conjecture from the underlying theory or from more fundamental properties of quantum gravity such as Black Holes (BHs) \cite{DGZ}, it has been tested in various cases in string theory~\cite{Grimm1}-\cite{Blum}. Such considerations   lead to similar bounds where the constant $c$ depend on $V$~\cite{Gia1,DGZ,TR}. At any rate, at the weak coupling and in the known string theory constructions,   $c$ turns out to be ${\cal O}(1)$  \cite{SV}.

In the same framework, another conjecture, the  Swampland Distance Conjecture (SDC) \cite{dSC2},
has been recently proposed. It is  motivated by the search of  consistent Effective Field Theories (EFTs) that do not belong to the Swampland and by the requirement that no new light degrees of freedom appear at low energy in the EFT. 
Indeed,  the SDC is related to changes in the moduli space of string vacua such that  a change along the geodesics in the field space is accompanied by the appearance of towers of light states with masses 
\begin{eqnarray}
m(\phi)\sim m_0 \, e^{-a|\Delta\phi|/{\mp}},   ~~~a\sim {\cal{O}}(1)>0,  \label{m}
\end{eqnarray}
descending from the UV for  trans-Planckian changes $|\Delta\phi|$ \cite{OV,KP,BP}.
%
In this short note we provide  evidence for the SDC by employing BHs in de Sitter space. In addition, we derive it by using  entropy bounds for the energy that can be stored in a region of spacetime. 

 
\section{The SDC from BHs}
In the presence of $N_s$ number of species, the effective cutoff $\Lambda_c$ 
reduces and it turns out to be \cite{Adler}-\cite{DLL}
\begin{eqnarray}
\Lambda_c=\frac{\mp}{\sqrt{N_s}}. \label{Lc}
\end{eqnarray}
As the distance  $\Delta \phi$ is increasing,  more and more states are becoming light and drop below $\Lambda_c$. In particular, 
 we have 
\begin{eqnarray}
 N_s\approx \frac{\Lambda_c}{m(\phi)}\approx\frac{\mp^{2/3}}{m^{2/3}(\phi)}.\label{ms}
 \end{eqnarray} 
 Therefore, with $m_0\sim \mp$,  we obtain 
\begin{eqnarray}
 N_s\approx e^{2a|\Delta \phi|/3\mp}.  \label{nns}
 \end{eqnarray} 
Consider  now a BH in a (quasi) de Sitter environment. The apparent horizon $R_s$ of such a  BH should be  inside the cosmic event horizon so that 
\begin{eqnarray}
R_s<H^{-1}, \label{Rsh}
\end{eqnarray}
or, since 
\begin{eqnarray}
R_s=\frac{M_\bh}{\mp^2}, \label{rs}
\end{eqnarray}
where $M_{\bh}$ is the BH mass,
\begin{eqnarray}
M_\bh<\frac{\mp^2}{H}. \label{c1}
\end{eqnarray}
In addition, $R_s$ should be larger than the cutoff length scale 
$\Lambda_c^{-1}$, i.e., 
\begin{eqnarray}
R_s>\Lambda_c^{-1},
\end{eqnarray}
or in other words, in the presence of many species, $M_\bh$ should satisfy the bound 
\begin{eqnarray}
M_\bh>\sqrt{N_s} \mp.  \label{c2}
\end{eqnarray}
Therefore, taking together Eqs. (\ref{nns}),(\ref{c1}) and (\ref{c2}), we get 
the bound 
\begin{eqnarray}
e^{a|\Delta \phi|/3\mp}<\frac{\mp}{H}. \label{h1}
\end{eqnarray}
The above  condition is necessary  for the BHs not to spoil the the effective description and of course it has to be true at any time. 
In particular, it should hold at the beginning of inflation where $H\sim \mp$ which is the common initial condition for large field models. Therefore, we get that the at the onset of inflation, the field should not vary more that 
\begin{eqnarray}
 |\Delta\phi|\lesssim \mp,
 \end{eqnarray} 
which is the SDC requirement.

 \section{The SDC from entropy bounds}
 Although the computation of the entropy of the tower of light states for trans-Planckian distances  in  a (quasi) dS spacetime is not possible, we may consider the case where  the  theory at low energies  consists of $N_s$ particle species. We will assume that such a system will not collapse to form a BH by supposing that the gravitational interactions between the particles can be neglected. 
The TCC can then be deduced by using the Bekenstein bound \cite{B1} on the entropy $S$ 
in a region of radius $R$ 
\begin{eqnarray}
 S<{\mpp} R^2. 
 \end{eqnarray} 
 Consider a spherical region of radius $R$  with $N_s$ particles at temperature $T$
 much larger their mass $m$, $T\gg m$. Since the particles are relativistic, the energy inside the sphere is 
\begin{eqnarray}
 E=\frac{4\pi^3}{45} N_s R^3 T^4,
 \end{eqnarray} 
 and their corresponding entropy is 
 \begin{eqnarray}
 S(R)=\frac{16\pi^3}{135} N_s R^3 T^3.  \label{s0}
 \end{eqnarray}
 The radius $R$ should be larger that the corresponding Schwarzschild radius $R_s=E/{\mpp}$ of a BH with the same energy $E$, so that 
 \begin{eqnarray}
 R\lesssim R_m\equiv\frac{{\mp}}{\sqrt{N_s} T^2}
 \end{eqnarray} 
Therefore, the maximum entropy in the sphere is 
\begin{eqnarray}
 S_{\rm max}\approx N_s R_m^3 T^3< {\mpp} R_m^2.
 \end{eqnarray} 
 This leads to the bound \cite{SPB2}
 \begin{eqnarray}
 T\lesssim \frac{{\mp}}{\sqrt{N_s}}, 
 \end{eqnarray}
 so that the maximum temperature is 
 \begin{eqnarray}
 T_m\approx \frac{{\mp}}{\sqrt{N_s}}.
 \end{eqnarray}
 Note that the connection between entropy bounds and maximal temperature is not new \cite{Bu1,Bu2}. 
%
By using now Eq. (\ref{nns}) we find that 
the maximum temperature is 
\begin{eqnarray}
T_m={\mp}  e^{-a|\Delta \phi|/{\mp}},
\end{eqnarray}
so that the condition $H<T_m$ leads again to the Eq. (\ref{h1})
and the subsequence considerations.

The same bound is obtained if one uses instead the Hubble entropy bound 
\cite{Ven2,Brus}
\begin{eqnarray}
S(R)< R^3 H {\mpp},
\end{eqnarray}
for the entropy in a region of linear size $R$. Using (\ref{s0}), we find that 
the maximum temperature is 
\begin{eqnarray}
T_m\sim \left(\frac{H {\mpp}}{N_s}\right)^{1/3}.
\end{eqnarray}
Then, the  condition $H<T_m$ leads again to  Eq. (\ref{h1}).  
%



\vskip .4in 
\noindent
{\bf \large{Acknowledgment}}
\vskip.1in
\noindent
 We thank V. De Luca, G. Dvali, G. Franciolini and D. L\"ust for several discussions and R. Brandenberger, P. Draper and A. Hebecker for useful correspondence.  A.R. is supported by the Swiss National Science Foundation (SNSF), project The Non-Gaussian Universe and Cosmological Symmetries, project number: 200020-178787. The work of A.K. is partially supported by the 
EDEIL-NTUA/67108600 project.


\begin{thebibliography}{99}

\bibitem{lrr} D.~H.~Lyth and A.~Riotto,
  Phys.\ Rept.\  {\bf 314}, 1 (1999)
  \arXivold{hep-ph/9807278}.

\bibitem{OOSV} 
G.~Obied, H.~Ooguri, L.~Spodyneiko and C.~Vafa,
\arXiv{1806.08362}{hep-th}.

\bibitem{BV} 
  T.~D.~Brennan, F.~Carta and C.~Vafa,
  PoS TASI {\bf 2017}, 015 (2017)
  \arXiv{1711.00864}{hep-th}.

\bibitem{Palti} 
  E.~Palti,
  Fortsch.\ Phys.\  {\bf 67}, no. 6, 1900037 (2019)
  \arXiv{1903.06239}{hep-th}.

\bibitem{V1} 
  C.~Vafa,
  \arXivold{hep-th/0509212}.  
  
\bibitem{OV} 
  H.~Ooguri and C.~Vafa,
  Nucl.\ Phys.\ B {\bf 766}, 21 (2007)
  \arXivold{hep-th/0605264}.  
  
  \bibitem{g1} G.~Dvali and C.~Gomez,
  JCAP {\bf 1401}, 023 (2014)
  \arXiv{1312.4795}{hep-th}.
  
  
  \bibitem{g2} G.~Dvali and C.~Gomez,
  Annalen Phys.\  {\bf 528}, 68 (2016)
  \arXiv{1412.8077}{hep-th}.
  
  \bibitem{g3} G.~Dvali, C.~Gomez and S.~Zell,
  JCAP {\bf 1706}, 028 (2017)
  \arXiv{1701.08776}{hep-th}.
  
  
  
\bibitem{dSC2} 
S.~K.~Garg and C.~Krishnan,
\arXiv{1807.05193}{hep-th}.

\bibitem{Kehagias:2018uem}
  A.~Kehagias and A.~Riotto,
  Fortsch.\ Phys.\  {\bf 66} (2018) no.10,  1800052
  \arXiv{1807.05445}{hep-th}.
  
\bibitem{Gia1} 
  G.~Dvali and C.~Gomez,
  Fortsch.\ Phys.\  {\bf 67}, no. 1-2, 1800092 (2019)
  \arXiv{1806.10877}{hep-th}.



\bibitem{dSC3} 
D.~Andriot,
Phys.\ Lett.\ B {\bf 785}, 570 (2018)
\arXiv{1806.10999}{hep-th}.


\bibitem{Cecotti:2018ufg} 
  S.~Cecotti and C.~Vafa,
  \arXiv{1808.03483}{hep-th}.

\bibitem{Ooguri:2018wrx} 
  H.~Ooguri, E.~Palti, G.~Shiu and C.~Vafa,
  Phys.\ Lett.\ B {\bf 788}, 180 (2019)
  \arXiv{1810.05506}{hep-th}.


\bibitem{dSfromSDC} 
A.~Hebecker and T.~Wrase,
Fortsch.\ Phys.\  {\bf 67}, no. 1-2, 1800097 (2019)
\arXiv{1810.08182}{hep-th}.

\bibitem{Garg:2018zdg} 
  S.~K.~Garg, C.~Krishnan and M.~Zaid Zaz,
  JHEP {\bf 1903}, 029 (2019)
  \arXiv{1810.09406}.

\bibitem{dSC4} 
D.~Andriot and C.~Roupec,
Fortsch.\ Phys.\  {\bf 67}, no. 1-2, 1800105 (2019)
\arXiv{1811.08889}{hep-th}.



\bibitem{Grimm1} 
  T.~W.~Grimm, C.~Li and E.~Palti,
  JHEP {\bf 1903}, 016 (2019)
  \arXiv{1811.02571}{hep-th}.  

\bibitem{Klaewer:2018yxi} 
  D.~Klaewer, D.~L\"ust and E.~Palti,
  Fortsch.\ Phys.\  {\bf 67}, no. 1-2, 1800102 (2019)
  \arXiv{1811.07908}{hep-th}.
  
 \bibitem{Corv} 
  P.~Corvilain, T.~W.~Grimm and I.~Valenzuela,
  JHEP {\bf 1908}, 075 (2019)
  \arXiv{1812.07548}{hep-th}.
  
  \bibitem{Lee:2019tst} 
  S.~J.~Lee, W.~Lerche and T.~Weigand,
  JHEP {\bf 1908}, 104 (2019)
  \arXiv{1901.08065}{hep-th}; 
  \arXiv{1904.06344}{hep-th};  
  \arXiv{1910.01135}{hep-th}.
  
\bibitem{Joshi:2019nzi} 
  A.~Joshi and A.~Klemm,
  JHEP {\bf 1908}, 086 (2019)
  \arXiv{1903.00596}{hep-th}.  
  
\bibitem{Heckman:2019bzm} 
  J.~J.~Heckman and C.~Vafa,
  Phys.\ Lett.\ B {\bf 798}, 135004 (2019)
  \arXiv{1905.06342}{hep-th}.  
  
  \bibitem{Lust:2019zwm} 
  D.~L\"ust, E.~Palti and C.~Vafa,
  Phys.\ Lett.\ B {\bf 797}, 134867 (2019)
  \arXiv{1906.05225}{hep-th}.
 
  
  \bibitem{Kehagias:2019akr} 
  A.~Kehagias, D.~L\"ust and S.~L\"ust,
  \arXiv{1910.00453}{hep-th}.
  
\bibitem{Grimm:2019ixq} 
  T.~W.~Grimm, C.~Li and I.~Valenzuela,
  \arXiv{1910.09549}{hep-th}.  
  
  \bibitem{Blum} 
  R.~Blumenhagen, M.~Brinkmann and A.~Makridou,
  \arXiv{1910.10185}{hep-th}.

\bibitem{DGZ} 
  G.~Dvali, C.~Gomez and S.~Zell,
  Fortsch.\ Phys.\  {\bf 67}, no. 1-2, 1800094 (2019)
  \arXiv{1810.11002}{hep-th}.

\bibitem{TR} 
  T.~Rudelius,
  JCAP {\bf 1908}, 009 (2019)
  \arXiv{1905.05198}{hep-th}.

\bibitem{SV} 
  P.~Agrawal, G.~Obied, P.~J.~Steinhardt and C.~Vafa,
  Phys.\ Lett.\ B {\bf 784}, 271 (2018)
  \arXiv{1806.09718}{hep-th}.


%
%
%
%
%
%
%
%
%
%
%
%
%
%
%

\bibitem{KP} 
  D.~Klaewer and E.~Palti,
  JHEP {\bf 1701}, 088 (2017)
  \arXiv{1610.00010}{hep-th}.


\bibitem{BP}
  F.~Baume and E.~Palti,
  JHEP {\bf 1608}, 043 (2016)
  \arXiv{1602.06517}{hep-th}.


\bibitem{Adler} 
  S.~L.~Adler,
  Phys.\ Rev.\ Lett.\  {\bf 44}, 1567 (1980); Phys.\ Lett.\  {\bf 95B}, 241 (1980).
 
 \bibitem{Zee:1981mk} 
  A.~Zee,
  Phys.\ Rev.\ Lett.\  {\bf 48}, 295 (1982).



%


\bibitem{Ven} 
  G.~Veneziano,
  JHEP {\bf 0206}, 051 (2002)
  \arXivold{hep-th/0110129}.



\bibitem{N1} 
  N.~Arkani-Hamed, S.~Dimopoulos and S.~Kachru,
  \arXivold{hep-th/0501082}.

\bibitem{N2} 
  J.~Distler and U.~Varadarajan,
  \arXivold{hep-th/0507090}.

\bibitem{N3} 
  S.~Dimopoulos, S.~Kachru, J.~McGreevy and J.~G.~Wacker,
  JCAP {\bf 0808}, 003 (2008)
  \arXivold{hep-th/0507205}.



\bibitem{SPB1} 
G.~Dvali,
Fortsch.\ Phys.\  {\bf 58}, 528 (2010)
\arXiv{0706.2050}{hep-th}.

\bibitem{SPB2} 
G.~Dvali and M.~Redi,
Phys.\ Rev.\ D {\bf 77}, 045027 (2008)
\arXiv{0710.4344}{hep-th}.


\bibitem{DLL}
G.~Dvali and D.~L\"ust,
  JHEP {\bf 0806}, 047 (2008)
  \arXiv{0801.1287}{hep-th}.

\bibitem{B1} 
 J.~D.~Bekenstein,
  Phys.\ Rev.\ D {\bf 23}, 287 (1981); 
  Phys.\ Rev.\ D {\bf 49}, 1912 (1994)
  \arXivold{gr-qc/9307035}.

\bibitem{Bu1} 
  R.~Brustein and G.~Veneziano,
  Phys.\ Rev.\ Lett.\  {\bf 84}, 5695 (2000)
  \arXivold{hep-th/9912055}.

\bibitem{Bu2} 
  R.~Brustein, S.~Foffa and G.~Veneziano,
  Phys.\ Lett.\ B {\bf 507}, 270 (2001)
  \arXivold{hep-th/0101083}.


\bibitem{Ven2} 
  G.~Veneziano,
  \arXivold{hep-th/9907012}.  
  
  \bibitem{Brus} 
  R.~Brustein,
  Lect.\ Notes Phys.\  {\bf 737}, 619 (2008)
  \arXivold{hep-th/0702108}.






%
%
%
%
%
%
%
%
%
%
%
  
 
 \end{thebibliography}
 \end{document}